# DESIGN OF AN INTEGRATED ANALYTICS PLATFORM FOR HEALTHCARE ASSESSMENT CENTERED ON THE EPISODE OF CARE

Douglas Teodoro[1], Nils Rotgans[2], Lucas Oliveira[3], Lilian Correia[4]

[1]Philips Research Brazil, São Paulo, Brazil
[2] Philips Design, Eindhoven, Netherlands
[3]Philips Research North America, Cambridge, USA
[4] Hospital Samaritano, São Paulo, Brazil

**Abstract**: Assessing care quality and performance is essential to improve healthcare processes and population health management. However, due to bad system design and lack of access to required data, this assessment is often delayed or not done at all. The goal of our research is to investigate an advanced analytics platform that enables healthcare quality and performance assessment. We used a user-centered design approach to identify the system requirements and have the concept of episode of care as the building block of information for a key performance indicator analytics system. We implemented architecture and interface prototypes, and performed a usability test with hospital users with managerial roles. The results show that by using user-centered design we created an analytical platform that provides a holistic and integrated view of the clinical, financial and operational aspects of the institution. Our encouraging results warrant further studies to understand other aspects of usability.

***Resumo:*** *A avaliação da qualidade e do desempenho do cuidado em saúde é essencial para melhorar os processos de cuidado e a gestão de saúde da população. No entanto, devido a falhas no design de sistemas de análise e a falta de acesso ao dados necessários, esta avaliação é frequentemente adiada ou não feita. O objetivo dessa pesquisa é investigar uma plataforma analítica avançada que permite avaliar a qualidade e o desempenho do cuidado em saúde. Nós utilizamos uma abordagem de design centrado no usuário para identificar os requisitos do sistema de análise, tendo o conceito de episódio de cuidado como o bloco de construção de informações para um sistema de análise de indicadores chave de desempenho. Nós implementamos os protótipos da arquitetura e da interface, e realizamos um teste de usabilidade com usuários com funções gerenciais em um ambiente hospitalar. Os resultados mostram que, usando um design centrado no usuário, fomos capazes de criar um conceito de análise que fornece uma visão holística e integrada dos aspectos clínicos, financeiros e operacionais da instituição. Esses resultados encorajadores justificam mais estudos para compreender outros aspectos de usabilidade.*

***Palavras-chaves****: Mineração de Dados; Qualidade do Cuidado; Banco de Dados*

## Introduction

In the past decade, Electronic Medical Records (EMRs) have evolved from simple patient health information management systems, where demographics, clinical and administrative information were stored and retrieved for patients, to become complex enterprise resource planning systems that operate over the entire healthcare system and include related services, such as administration, finance, supplier management, human resources, and decision support. They also provide subsidies for billing and reimbursement of service providers, and serve as a base for organizational support and cost management of healthcare facilities[1]. Thus, nowadays there are several examples of non-medical functionalities that have been integrated into EMRs and are considered essential by many healthcare enterprises. EMRs also provide academic functionality to support clinical research, epidemiological studies and information sharing between multi-professional teams.

The availability of a large amount of patient records in the EMR, together with administrative, operational, and financial information, enables integrated key performance indicator (KPI) analytics across the hospital's patient population. KPI analytics provide a means to monitor and assess clinical effectiveness, patient safety, efficiency, staff orientation, and governance for quality improvement in the healthcare settings[2]. Using metrics provided by KPIs, decisions can be made in the enterprise to improve clinical, operational and financial management[3]. For example, when occupation rate data is available to decision makers in a timely manner, actions can be taken on the agenda of a given period to increase or decrease the number of booked patients or staff according to the target hospital occupation metrics. Similarly, if the time between sepsis identification and antibiotic administration is known, actions can be taken to reduce potential delays in the treatment and increase survival rates[4].

However, it is often the case that EMR systems are composed of silos containing heterogeneous clinical, administrative, operational and financial information spread in several modules or subsystems. Amongst other problems, the lack of a central and integrated data repository delays the computation of KPI metrics, jeopardizing their utilization as an actionable information source[5]. It is common that KPIs, such as contribution margin, are consolidated at the end of relatively long periods, not allowing corrective actions to be taken within the given analysis period. Furthermore, KPIs may not provide comprehensive assessment of institution metrics as they are usually calculated without considering the context of the whole healthcare enterprise, using sparse and fragmented pieces of information (e.g., using only financial or only clinical data) that are not comprehensive enough to reveal important insights for the institution.

To improve performance of population health management and the quality of health care, in this work we investigate the use of an integrated analytics repository composed of several data sources to seamlessly capture all the events related to a patient's episode of care and provide near-real time information for hospital management decision making. User insights generation and co-design sessions were organized to understand daily issues that healthcare data analysts face during hospital quality and performance assessment. Using the episode of care (EoC), i.e., the set of events related to a patient treatment, including clinical but also financial and administrative data, as a crucial backbone of information, we build a prototype framework for characterizing patient cohorts, extracting KPIs and forecasting hospital performance indicators to provide results that are more accurate and informative than those based on information silos. The prototype concepts were validated using usability workshops, where the requirements were assessed. In this paper, we introduce the preliminary analytics architecture and the results of the user evaluation.

**Methods**

**Study Context**

In this study, we use retrospective anonymized data from Philips Tasy, an EMR system with more than 80 modules that manages clinical, administrative, operational and financial healthcare information. The dataset was provided by Hospital Samaritano – São Paulo, a 300 bed philanthropic hospital. The data was de-identified using the HIPAA guidelines and permission to use it was granted by both Hospital Samaritano and Philips Research internal ethics committees.

**Requirements Elicitation**

To deepen our understanding about the issues that healthcare data analysts and decision makers face during hospital quality and performance assessment, user insights generation and co-design sessions were organized. The methodological approach was based on participatory design[6], where users are seen as experts in their own experience, and projective and constructive exercises, such as collages, also coined as "visual literacy", are used to support users sharing their experiences and reflecting on them in deeper ways. The outcomes of these sessions define the main issues that users are facing and possible solutions to solve them.

The sessions were held at Hospital Samaritano between 30/03/2015 and 02/04/2015. Fifteen members from four areas of the hospital – clinical, epidemiological, financial and operational – involved with data analytics activities participated in four co-design sessions of approximately 180 minutes each (1 session per area). In addition to these sessions, three observation sessions one hour each were conducted with the users to observe their daily analytics activities and interactions with current tools and solutions. Finally, a validation session was held to confirm and enrich the information gathered in the previous sessions, which took 120 minutes.

**Prototyping Methods**

In this section, we describe the methodology used to create the platform prototypes. The architecture design followed the Service Oriented Architecture (SOA) approach, where each KPI was represented by a different Representational State Transfer (REST) service[7]. The backend services were created using JAVA and Scala languages. The database choice was Hive, a high availability NoSQL database[8]. For the frontend interface, two prototypes were created. The first was a design mockup built using Adobe Illustrator, for concept creation, and Microsoft PowerPoint, to enable dynamic navigation through the design concepts using links. The second prototype was a runnable software, implemented using HTML5 and JavaScript technologies.

**User Validation**

To validate the prototypes, we organized evaluation sessions with the decision makers at Hospital Samaritano – São Paulo. In total, seven sessions of 1.5 hours each were held with seven managers from clinical and financial domains (3 and 4 users respectively). The goal of these sessions was to understand whether the prototypes created met the user analysis needs and to gather feedback on what further needs they might have so that the prototypes can be continuously improved in an agile fashion. The decision to focus on financial and clinical domains was based on analysis priorities defined by the hospital. The rationale to restrict the evaluation to only two analysis domains was to simplify the study. In the validation sessions, we performed moderated usability tasks combined with specific questions about how each feature supports users' needs and what could be done to improve them. During the tasks, users were encouraged to talk while they were using the prototype.

**Results and Discussion**

**Identification of Requirements**

Despite coming from different areas/departments of the hospital, after the four co-design sessions, a set of issues and needs appeared as a pattern when analyzing KPIs at the hospital. We learned that users strive to gain as much insight as possible from the data. They want to make data actionable to create plans and activities that will bring improved results. However, this is a challenge since the preparation of the data (collection and consolidation) takes too much time, leaving no space for proper analysis and consecutive actions. In their view, the approximate percentage of time that it takes them to prepare the data is 80%, leaving only 20% to analyze it. Ideally, they would like to invert those percentages so that can they spend more time (80%) analyzing and executing actions to improve current issues, and only 20% preparing the data. Table 1 resumes the main findings of the data requirements workshop.

Table 1: Needs and solutions for healthcare data analytics.

| Needs and constraints | Solutions |
| --- | --- |
| Cumbersome and time consuming data collection and consolidation, limiting the time to generate insights | Centralized (impartial) location for the data |
| Lack of integration and synchronization between the hospital areas and the information | Use episode of care as a holistic view for operational and clinical activities |
| Lack of context of the data to drill down roots causes | Feature to track down targets by disaggregating dataset attributes |
| Data is delayed, therefore by the time the issue is found it is already late | (Near-)Real time data and easy ways to create thorough correlations combing clinical and operational attributes |
| Lack of ways to do forecasting and simulations | Feature to create predictive scenarios and assess possibilities in easy and fast ways |

Users argue that the use of delayed data and retrospective analyses are limiting factors. Their preference is to have (near-) real time data and analyses, and predictive and simulating tools that would allow them to understand issues as they happen and to promptly act upon them. For example, a study participant said during the operational group session that: *"I aggregate information to make my analysis. But it only happens one week after the end of the month. Meanwhile, I go into the situation blind... Will I meet my goal or not? I would like to have this information daily. How much I have produced, how much I will produce. From the budget point of view, am I going to make it or not? This information only appears in a reliable form after any time for action has past. Then, I can only sit and regret"*.

Additionally, the lack of centralized data repository is a relevant issue during the user's data analytics activities. Data needs to be fetched from different applications and placed into a new format to be queried. For example, in one of the sessions, the moderator asked *what is the most critical situation in the analytics process*, and the users replied *the manual data collection and integration from different systems*. Moreover, the lack of one integral view is an issue for data analytics. Users argue that they have a fragmented view of any problem that they analyze. This is due to the lack of context or connection between the different data locations (administration, operational, supplies, etc.). The following passage stated during the financial group session illustrates this: *"The whole [institutional view] is difficult to bring together because it [the information] does not converge... Today, we do not have this integrated view of the patient, which would look at the patient as whole, from the moment they arrive until the moment they leave [the institution]"*.

**Architecture and Interface Prototypes**

**Architecture**

Based on the requirements identified with the end users, we proposed the architecture for healthcare quality and performance assessment presented in Figure 1. This architecture provides the modules for extracting and integrating all relevant data needed to create aggregated, non-fragmented, and context-aware KPI views. It allows the integration of various healthcare data sources to provide a comprehensive understanding of patient population flows within the hospital and the metrics associated with them. The key element of the architecture is a central and integrated EoC repository, which captures all events related to a patient treatment, including clinical but also financial and administrative data, as a single piece of information. By connecting the KPI metrics to the EoCs, at the population level, the system allows the information to be better contextualized and thus understood. We use a private cloud platform, based on the Hadoop ecosystem, to guarantee the architecture scalability in terms of both storage and computing power.

The proposed architecture is composed of five different modules: EoC Repository, EoC Extractor, EoC Builder, EoC Classifier and the KPI Processor. The EoC Extractor module provides technical and syntactic interoperability to the analytics platform[9]. The data coming from multiple data sources are highly heterogeneous, with different data types, data models, formats and semantics. First, this module is responsible for interfacing the different data sources, homogenizing APIs and connection protocols, to extract events associated to the episode of care. Second, it converts the different data models into a single format, a document model using the JSON syntax[10]. Finally, it provides a connector to the episode of care repository, allowing data to be loaded into it. Data streams are routinely loaded into the central repository using time stamps of the source datasets.

The EoC Repository stores all information related to a patient's episode of care found within the healthcare institution. It aggregates the data from several healthcare data sources to create a unified register with the patient population flow, encoded in the episodes of care. Due to the document-based nature of the episode of care, this repository is backed by a NoSQL database, providing high model flexibility and retrieval performance[8].

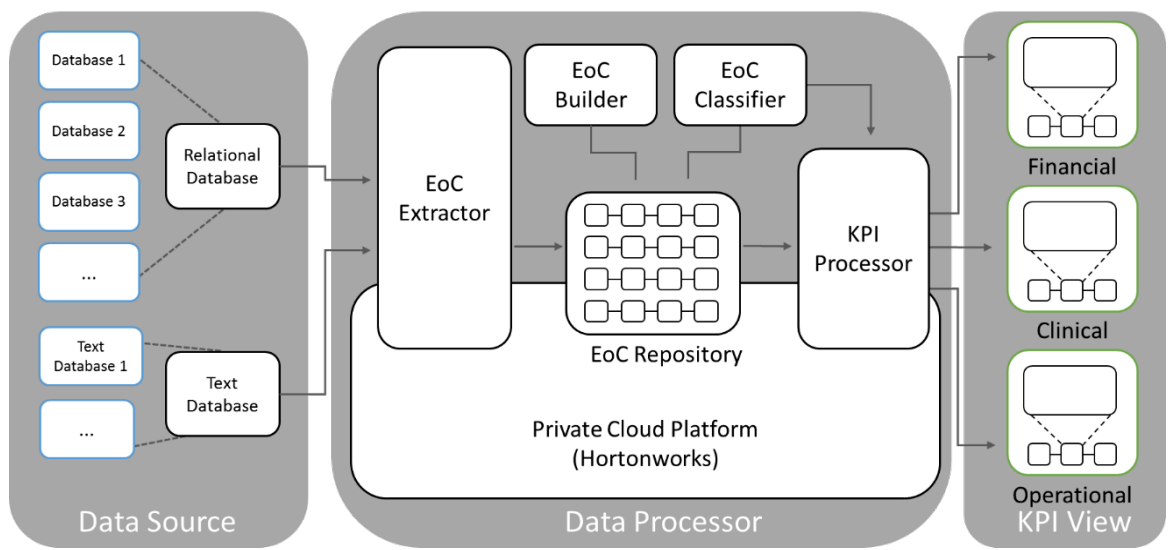

Figure 1: Integrated KPI Analytics Architecture. EoC - Episode of Care.

The *EoC Builder* module is responsible for linking the different events belonging to a patient's episode of care and constructing an array structure that stores all this information in the *EoC Repository*[11]. The *EoC Classifier* is an auxiliary module, whose function is to automatically create episode of care cohorts so that KPIs can be associated to certain patient groups, and root-cause analyses can be performed[12]. Finally, the *KPI Processor* provides means to actually calculate the KPIs. This module implements the statistical functions to compute the different KPI metrics displayed in the user interface, and provides a REST API to retrieve these data based on some parameters, such as period (e.g., from and to date), type of analyses (mortality, length of stay, etc.), and stratification groups (e.g., gender and age). It accesses to the *EoC Repository* online (near real-time) and computes the statistics based on the query parameters.

**Interface Prototype**

The interface prototypes implement the KPI View modules of Figure 2. This interface was used to validate user insights providing the end user a way to navigate through the healthcare KPIs, drilldown a specific information of interest and give some feedback about their experience. The features implemented in the interface realize the requirements identified during the workshop (Table 1). In particular, it was designed to provide an integrated view of the institution, having the patient (population) in the core of the analyses. The interface was created using user centered design methodology and implemented using HTML5. Figure 2a shows the high-level overview of the financial KPIs while Figure 2b shows KPIs for stent procedure in the cardiology department for patients with a length of stay of 7 days or more.

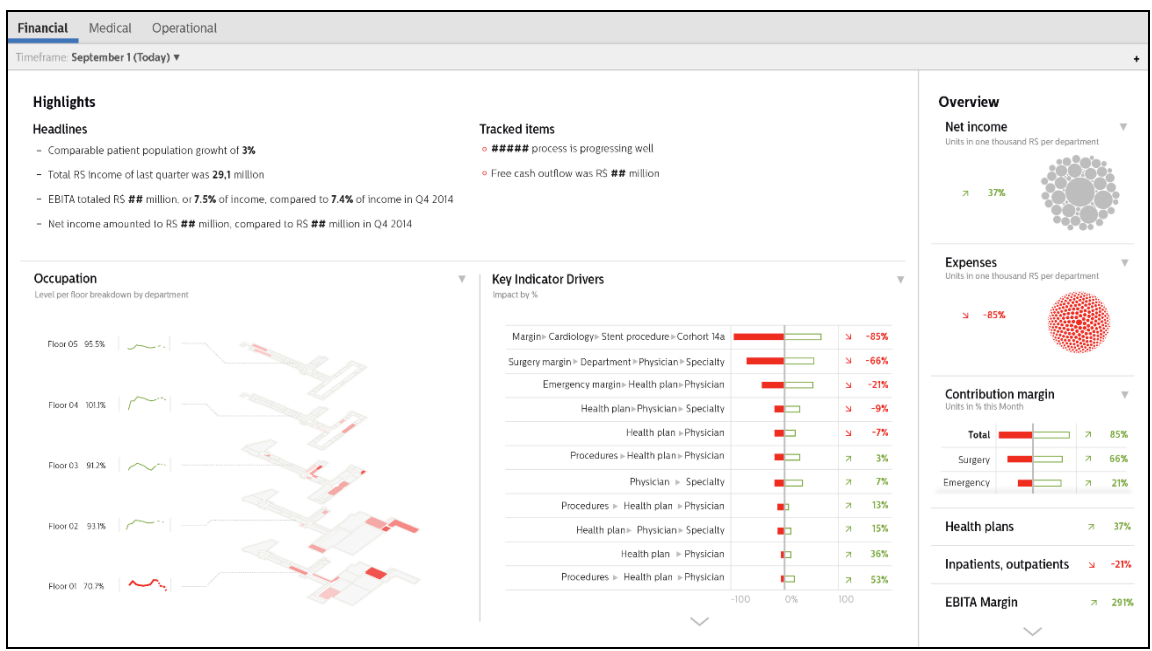

(a)

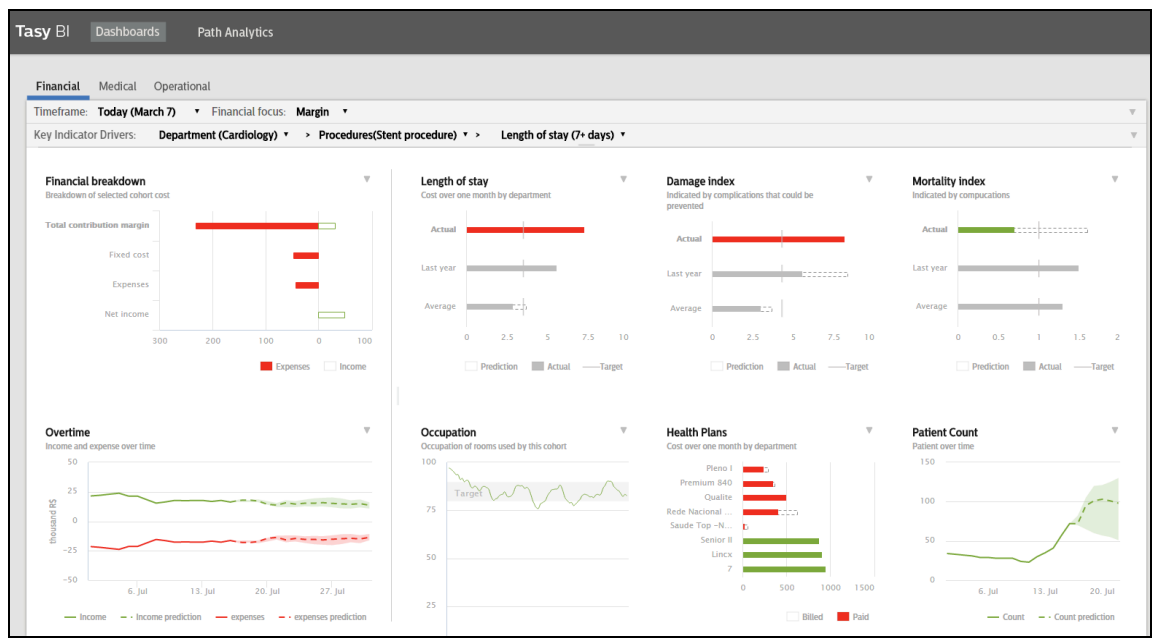

(b)

Figure 2: Preliminary user interface implementing few KPIs.

**User Validation**

The prototype validation was performed during seven different 1.5 hour sessions with decision makers from clinical and financial domains (3 and 4 users respectively). Usability tasks were performed combined with specific questions on how each feature supports users' needs, what could be done to improve them and what the rationale is behind their feature priorities. A list of findings regarding these sections are summarized in Table 2.

Users found the concept and idea of having all data in a single system valuable. The possibility of reviewing data through charts was found useful and considered an important feature needed for improving their work. Navigation and filters were the main problems that need adjustment to improve the overall experience of the system. Users expressed the need for creation of user-customized dashboards, however they also said that changing an existing dashboard, e.g., adding a new KPI or requesting a new filter, is not a usual task for them. Finally, they found the ability to interact with the dashboard important in order to be able drill down the data for deeper analysis.

Table 2: Findings of the usability validation sessions.

| User | Worked Well | Needs Improvement | Improvements |
|---|---|---|---|
| Clinical and Financial | Having important information displayed as graphics | Reading some design elements (titles and chart axes) was difficult | Using darker grey or even black will result in improved readability |
| Clinical | Chart designs were considered easy to read and understand Feature to compare a cohort to another and to | Users did not understand the filter criteria The options to modify a KPI chart (arrow) was | Filter providing a combination of tokens and dropdowns Moving the arrow closer to the title will |
|  | the hospital's average | not easily found by most users | make it more visible |
|  | Filters were considered helpful to navigate through detailed information with different points of view | Non-standard charts were usually considered difficult to understand by users | Using standard stacked bar chart with legend |
| Financial | Having a dashboard with rich information available at once | Difficulties finding the down arrow to see more indicators | A link or button labeled "More indicators" will be clearer to users |
|  | Track items was a useful feature | Took some time to notice some design elements and features (headlines and tracked items) | Consider using design elements that draw users' attention |
|  | Value to see projections of what is expected for the coming months | Solid and outlined bars confused the users | Consider having only solid bars |

**Conclusion**

By linking the different KPI views to the essential healthcare enterprise process (i.e., the episode of care), the architecture proposed here contextualizes the information generated by the KPIs and makes it uniform throughout different institution sectors. We have learned that users are interested in tools that extend their understanding of operational patterns of healthcare organizations. The way that the data is logically stored is less important for them. These tools should have customable interfaces that reflect the user workflow in which the data can intuitively be explored according to the context. An important aspect pointed out by users is the ability to interact with the dashboard and drill down to further analyze the data. Findings of this work will lead to new concepts that will be created and validated by users in our future work. An important aspect of this research is the understanding of the needs of different users and the customization of the developed tool to meet those needs.

Even though the number of users was relatively small and statistically significant quantitative data were not obtained, the insights provided will help in understanding the requirements by different users. The next steps involve the training and test algorithms for episode of care data linkage using several sources and episode of care classification, and the finalization of the prototype for a larger clinical usability assessment.

**Contact**

Douglas Teodoro
SIB Swiss Institute of Bioinformatics
Battelle campus
B Building - office #3.22
Rue de la Tambourine 17
CH-1227 Carouge
douglas.teodoro@sib.swiss
+41 22 546 3571